\documentclass[prl,aps,graphicx,float]{revtex4}
\usepackage{epsfig}
\usepackage{bm}

\begin{document}

\preprint{}

\newcommand{\stef}[2]{$\blacktriangleright${\sc round #1:} 
{\em #2}$\blacktriangleleft$}

\title{Nonequilibrium quantum phase transition in itinerant electron systems}

\author{D. E. Feldman}

\affiliation{Department of Physics, Brown University, 
Providence, RI 02912}


\begin{abstract}
We study the effect of the voltage bias on the ferromagnetic phase transition in a 
one-dimensional itinerant electron system.
The applied voltage drives the system into a nonequilibrium steady state with 
a non-zero 
electric current. The bias changes
the universality class of the second order ferromagnetic transition. 
While the equilibrium transition belongs to the universality class of the uniaxial 
ferroelectric, we find the mean-field behavior near the nonequilibrium 
critical point.

\end{abstract}


\maketitle

When external fields drive a many-body system into a nonequilibrium steady state, 
a parameter change may result in a transition between two nonequilibrium phases. 
Such nonequilibrium transitions were observed in 2D electron gases driven with 
microwave radiation \cite{zrs1,zrs2}, granular matter \cite{Aranson}, 
and many other systems. 
Scaling and universality hold near second order phase transitions both in 
and far from thermal equilibrium \cite{dds}. 
Hence, one can describe them with renormalization group. 
This helped to achieve significant progress in the theory of nonequilibrium 
classical phase transitions \cite{dds}. 

The possibility to achieve a high degree of isolation from environment for ultracold 
atoms \cite{coldatoms} opens a way to study quantum dynamics in many-body systems far 
from equilibrium. This has stimulated interest to nonequilibrium quantum phase 
transitions of strongly interacting bosons \cite{apdhl}. 
Another type of nonequilibrium 
quantum transitions is possible in itinerant electron systems driven into states with 
nonzero electric current. 
In this Letter, we address a nonequilibrium ferromagnetic transition 
of 1D itinerant fermions. Since this transition exists only at zero temperature,
it cannot be considered as a classical phenomenon, 
in contrast to the decay of supercurrents 
of interacting bosons that can be understood as a classical 
modulational instability \cite{apdhl}.

Equilibrium magnetic phase transitions of itinerant electrons have attracted much 
attention \cite{it,it1,it2,it3}. The quantum critical behavior at these 
transitions is 
a subtle and challenging problem due to the coexistence of two types of soft modes. 
In addition to the order parameter fluctuations, the theory must account for gapless 
fermions which induce effective long-range interactions between the order parameter 
fields in different points \cite{it}. Most studies have focused on itinerant systems 
in two and three dimensions. According to the Lieb and Mattis theorem \cite{LM} there 
is no ferromagnetism in a wide class of one-dimensional models with spin-independent 
interactions. Yet ferromagnetic quantum phase transitions were found in many 
one-dimensional itinerant systems both with and without magnetic anisotropy. 
In particular, recent numerical simulations \cite{sim} discovered a quantum critical 
point in a tight-binding model with next nearest neighbor hoping. It was also 
argued \cite{fm07} that the 0.7 anomaly in transport properties of quantum wires 
\cite{fm07,07} could be explained by spontaneous magnetization of itinerant electrons. 
Recently,
the equilibrium quantum phase transition of itinerant electrons in 1D was 
theoretically 
investigated in Refs. 
\onlinecite{1fm1,1fm2}.

In this Letter we study the effect of a voltage bias on the quantum ferromagnetic 
transition in a one-dimensional itinerant electron system. A low voltage cannot 
significantly affect the magnetization far from the transition. However,
it is a relevant perturbation in the critical domain and changes the universality 
class of the transition. While the equilibrium second order transition \cite{1fm2} 
belongs 
to the universality class of the uniaxial ferroelectric \cite{lkh,sm,feldman}, 
we find that the nonequilibrium critical behavior is mean-field like, {\it i.e.} 
the magnetization $m\sim\sqrt{r}$, where $r$ is the distance from the critical point. 

The paper is organized as follows: first we introduce the model, 
then briefly discuss the phase transition within the mean-field Stoner approach, 
and finally show that the mean-field approximation gives the exact critical behavior.

We consider the Tomonaga-Luttinger model with the Hamiltonian
\begin{equation} 
 \label{1} H=\int dx \bigg\{\sum_\sigma[   \psi_{\sigma R}^\dagger(x)
  \epsilon(-i\hbar\partial_x)\psi_{\sigma R}(x) + \psi_{\sigma L}^\dagger(x) 
\epsilon(-i\hbar\partial_x)\psi_{\sigma L}(x) ] 
  +\sum_{\sigma\sigma'}\int dy
  K_{\sigma\sigma'}(x-y)  \rho_{\sigma}(x) \rho_{\sigma'}(y) \bigg\}, 
\end{equation}
where $\psi_{\sigma R}^\dagger$ and $\psi_{\sigma L}^\dagger$ are the creation 
operators
for right- and left-moving electrons with the spin $z$-component $\sigma=\pm 1/2$, 
$\psi_{\sigma}^\dagger =
\psi^\dagger_{\sigma R}+\psi^\dagger_{\sigma L}$ gives the conventional electron
creation operator, $\rho_\sigma=\psi_\sigma^\dagger \psi_\sigma$ 
is the electron density, 
and $K_{\sigma\sigma'}(x-y)$ 
the interaction
strength. We assume that the long-range Coulomb interaction is screened by the gates 
so that $K_{\sigma\sigma'}(x-y)$ decreases rapidly for large $(x-y)$.  
The spin dependence of $K_{\sigma\sigma'}$ determines the symmetry of the model.
As in Ref. \onlinecite{1fm2} we will focus on the technically simplest case of the Ising 
symmetry $Z_2\times U(1)$
($U(1)$ is responsible for the conservation of $S_z$; $Z_2$ describes the symmetry 
with respect to the $\sigma\rightarrow-\sigma$ transformation). Note that the 
nonlinearity of the spectrum $\epsilon(k)$ is relevant near the critical point. 
As discussed in Ref. \onlinecite{1fm1} this nonlinearity is important for the very 
existence 
of the quantum critical point.

The equilibrium transition in the model (\ref{1}) was considered in Ref. \onlinecite{1fm2}.
We will  study what changes in the presence of a voltage bias. We will show that the non-equilibrium problem 
is equivalent to a certain equilibrium model which lacks the time-reversal symmetry. The absence of this symmetry
is related to the fact that the ``Fermi velocities'' of the right- and left-moving electrons are different.
The critical behavior of the latter model can be found exactly. We will show that the diagrams for the magnetization are free from infra-red divergencies. Hence, the system belongs to the mean-field universality class.

Before we develop a systematic approach let us briefly discuss the simplified 
Stoner picture. Let the number of the right/left-movers 
with spin $\sigma$ be $N_0+N_{R/L \sigma}$, where $N_0$ denotes 
the number of the states with the negative energy $\epsilon(k)$. 
The Stoner model is described by the Hamiltonian 
$H=\sum_\sigma\sum_{D=R,L} U(N_{D \sigma}) -
KM^2$, where
$U(n)=\int_{2\pi N_0}^{2\pi (N_0+n)}\epsilon(k) \frac{dk}{2\pi}$, 
and the magnetization  
$M=(N_{R [1/2] }+N_{L [1/2] }-N_{R [-1/2]}-N_{L [-1/2]})/2$. 
The numbers of the right- and left-movers are integrals of motion.
The chemical potentials  of the right- and left-movers $\mu_{R/L}=\pm eV/2$ are different.
Hence, at zero temperature the system is in the ground state of the effective Hamiltonian
$H'=H-\mu_R N_R - \mu_L N_L=H 
-V(N_{R [1/2]}+N_{R [-1/2]}-N_{L [1/2]}-N_{L [-1/2]})/2$.
From the energy minimum condition $\partial H'/\partial N_{D\sigma}=0$ we find that
the maximal energies $\epsilon(k)$ of the right-moving particles with the 
opposite spins 
differ by
$2 \Delta E= K\int_{-\Delta E}^{\Delta E}dE[\rho(E+V/2)+\rho(E-V/2)]$,
where $\rho(E)$ is the density of states. Expanding $\rho(E\pm V/2)$ in powers
of $E,V$ one finds $\Delta E[1-2K(\rho(0)+\rho''(0)V^2/8)]-K\Delta E^3\rho''(0)/3=0$. 
This equation describes a second order phase transition, if 
$\rho''(0)<0$. The voltage bias shifts the transition 
to the greater interaction strength
$K=K_c$, $\Delta K_c\sim V^2$. The magnetization exhibits the mean-field behavior $M\sim\Delta E\sim\sqrt{K-K_c}$. 
We will see below that the mean field prediction for the critical behavior of 
the magnetization is valid beyond the Stoner approximation. 

In the Stoner picture,
electrons move in a spin-dependent self-consistent field without backscattering. 
Hence, the electric current is $I=2e^2V/h$. The spin current $I_s=0$.

In a systematic approach,  we need to include the leads into the model to describe the voltage bias. 
We use the standard model \cite{leads,FG,TSDG, FSV} for the Fermi-liquid 
leads adiabatically connected to the wire.  We assume that the Hamiltonian (\ref{1}) 
is applicable for $|x|<L$ only, where $2L$ is the wire length. 
At large $|x|$ the interaction strength $K(x-y)$ is zero. 
We assume that at the initial moment of time $t=-\infty$ the interaction 
$K_{\sigma\sigma'}=0$. Thus, the numbers of the left- and right-movers are 
integrals of motion at $t=-\infty$. Hence, the initial state can be described 
in terms of two different chemical potentials $\mu_R=eV/2$ and $\mu_L=-eV/2$ for 
the right- and left-moving electrons, where $V$ is the voltage bias. 
At later times $t>-\infty$ the interaction is gradually turned on. 
It is convenient to switch to the interaction representation so that 
the Hamiltonian becomes $H'= H-\mu_L N_L -\mu_R N_R$, where $N_L$ and $N_R$ 
are the numbers of the left- and right-movers. The initial state is the ground 
state of $H'$ at $t=-\infty$. The interaction representation introduces time 
dependence into the electron creation and annihilation operators. 
Hence, the evolution of the system should be described with the 
nonequilibrium Keldysh formalism \cite{Keldysh,FG,TSDG,FSV}. 
The bare Keldysh Green functions are determined by the ground state 
of the Hamiltonian $H'$ at $t=-\infty$. We will see however, that the time-dependence drops out from 
the problem and a simpler equilibrium technique can be used.


The bosonization approach \cite{Giamarchi} allows one to reduce one-dimensional 
problems with strong electron interaction to models of weakly interacting bosons 
which can be studied perturbatively in the interaction strength. 
Following the standard bosonization procedure we define four Bose-fields 
$\phi_{\sigma R}$ and $\phi_{\sigma L}$ such that

\begin{equation}
\label{2}
\psi_{\sigma R}\sim\exp(i\phi_{\sigma R}+i k_{f R}x-iVt/2);
\psi_{\sigma L}\sim\exp(-i\phi_{\sigma L} - i k_{f L}x+iVt/2),
\end{equation}
where $k_{f R}$ and $k_{f L}$ play the role of the effective ``Fermi momenta'' 
and are proportional to the average electron density of the right- and left-movers, 
$\rho_{L/R}=\sum_\sigma\rho_{\sigma L/R}=(\sum_\sigma\partial_x\phi_{\sigma L/R}+
2k_{f L/R})/(2\pi)$, $\langle\rho_{L/R}\rangle=k_{f L/R}/\pi$. 
We omit Klein factors in Eqs. (\ref{2}) since they are not important 
for the following calculations. 
Two wave vectors $k_{f L/R}$ differ since the applied voltage bias leads 
to the different densities of the right- and left-moving electrons.  
In most cases this difference can be neglected since the voltage is much 
smaller than the bandwidth. However, it is important in our problem. 
The bosonized action contains two types of terms.
The first type is a product of derivatives of the Bose-fields with 
respect to $t$ and $x$ (``forward scattering'' terms). 
The second type includes ``backscattering'' terms proportional 
to $\exp(in[\phi_c/2+(k_{f R}+k_{f L})x-Vt]+il\phi_s/2)$, where 
the charge and spin fields 
$\phi_c=\sum_\sigma(\phi_{\sigma R}+\phi_{\sigma L})$ and 
$\phi_s=\sum_\sigma 2\sigma (\phi_{\sigma R}+\phi_{\sigma L})$ 
are related to the charge and spin densities as 
$\rho=e(\partial_x\phi_c+2k_{f L} + 2k_{f R})/(2\pi)$ and 
$m=\partial_x\phi_s/(4\pi)$;
$n$ and $l$ are integers, $(n+l)$ is even. 
The dual fields $\theta_c=\sum_\sigma(\phi_{\sigma R}-\phi_{\sigma L})$ and 
$\theta_s=\sum_\sigma 2\sigma (\phi_{\sigma R}-\phi_{\sigma L})$ do not 
appear in the exponents due to the conservation of the charge and the $z$-component 
of the magnetization.
The terms of the second type with $n\neq 0$ are suppressed by their rapidly 
oscillating coordinate dependence. Hence, as usual in bosonization they are 
not important for the low-energy physics \cite{Giamarchi} which determines 
the critical behavior. We will also see that the terms with $n=0,l\neq 0$ are 
irrelevant near the phase transition. Thus, only the terms of the first type 
should be taken into account. This means that the terms which should be included 
in the action do not depend on time explicitly. Hence, it becomes possible to 
use the equilibrium imaginary time formalism instead of the non-equilibrium 
Keldysh technique \cite{Keldysh,FG,TSDG,FSV}.

The Euclidean action contains four interacting chiral fields $\phi_{\sigma R/L}$:

\begin{equation}
\label{4}
L=\int dx d\tau \bigg\{\frac{1}{4\pi}\sum_\sigma[\partial_x\phi_{\sigma R}
(i\partial_\tau+v_R\partial_x)\phi_{\sigma R}
+\partial_x\phi_{\sigma L}(-i\partial_\tau+v_L\partial_x)\phi_{\sigma L}]+
\sum_{\sigma,\sigma'=\pm 1/2; D,D'=R,L}u_{\sigma\sigma' D D'}\partial_x\phi_{\sigma D}
\partial_x\phi_{\sigma' D'}+\dots\bigg\},
\end{equation}
where dots denote higher order terms. Note that the ``Fermi velocities'' 
$v_{R/L}=\frac{d\epsilon(k=k_{f R/L})}{dk}$ are different for the right- and 
left-movers. 
We rewrite the action in terms of $\phi_c,\phi_s,\theta_c,\theta_s$ and integrate 
out $\theta_s$ and $\theta_c$. The symmetry with respect to the magnetization 
reversal results in the spin-charge separation, {\it i.e.} the quadratic part of 
the resulting action is the sum of a contribution which depends on $\phi_c$ only and 
a contribution that depends on $\phi_s$ only:

\begin{equation}
\label{5}
L^{(2)}=L^{(2)}_s(\phi_s)+L^{(2)}_c(\phi_c).
\end{equation}
In general, the spin-charge separation does not extend beyond the terms quadratic 
in the Bose-fields. However, similar to Ref. \onlinecite{1fm2} the interaction between 
the spin and charge degrees of freedom is not important near the phase transition. 
One can show that this interaction does not change the critical behavior. 
Thus, we can focus on the
case when Eq. (\ref{5}) is valid in all orders and hence the magnetization depends 
only on the spin part of the action $L_s(\phi_s)$. A straightforward modification 
of our approach demonstrates that our results
are independent of this assumption (cf. Ref. \onlinecite{1fm2}). 


After an appropriate rescaling of the time variable, the action takes 
the following form:

\begin{equation}
\label{6}
L=\int d\tau dx \bigg\{\frac{1}{2}[(\partial_\tau\phi_s)^2+
2ic\partial_\tau\phi_s\partial_x\phi_s
+r(\partial_x\phi_s)^2]+\dots\bigg\}, 
\end{equation}
where the dots denote terms with higher powers of $\phi_s$ and/or gradients. 
The terms with an odd (even) number of time derivatives are imaginary (real). 
The action (\ref{6}) differs from the standard expression \cite{Giamarchi} 
due to the presence of the second imaginary term. Technically this term appears 
because two ``Fermi velocities'' $v_{ R}$ and $v_{L}$ are different. 
Its presence reflects the fact that the action 
(\ref{4}) lacks time-reversal symmetry. In most problems the second term 
can be neglected. Indeed, the constant $c$ is proportional to the voltage bias 
which is much smaller than the bandwidth. In particular, the second term is not 
important far from the transition since it is smaller than the sum of the first 
and third terms in (\ref{6}). However, at the critical point, $r$ vanishes 
and hence the second term is not negligible. 
At negative $r$ the quadratic part of the action (\ref{6}) becomes unstable. 
This signals a phase transition.  To make sure that the system is stable at 
negative $r$ one needs to take into account the contribution  of the form
$L_{2,2}=b(\partial_x^2\phi_s)^2$ and nonlinear terms such as 
$L_{1,4}=u(\partial_x\phi_s)^4$ 
denoted by dots in Eq. (\ref{6}). $L_{2,2}$ emerges from 
the coordinate dependence of the 
interaction $K(x-y)$, Eq. (\ref{1}), \cite{1fm2}. 
$L_{1,4}$ describes the nonlinearity of 
the spectrum $\epsilon(k)$ \cite{Haldane}.

In the critical point the quadratic part of the action is thus proportional to 
$(\partial_\tau\phi_s)^2+
2ic\partial_\tau\phi_s\partial_x\phi_s+b(\partial_x^2\phi_s)^2$. 
At each length scale $1/k$ the poles of the propagator determine 
two frequency scales 
$\omega_1\sim k$ and $\omega_2\sim k^3$. Hence, similar to other models  of itinerant 
electron systems \cite{it2},
the choice of scaling dimensions is subtle. 
Indeed, if the scaling dimension of the coordinate $[x]=-1$ then two different 
frequency scales correspond to $[t]=-1$ and $[t]=-3$. However, for any of these 
two choices, power counting 
yields a negative or zero scaling dimension for all possible non-linear operators. 
This suggests the mean-field behavior.
Below we show that the critical behavior is indeed mean-field-like without 
logarithmic corrections. 

\begin{figure}
   \epsfig{file=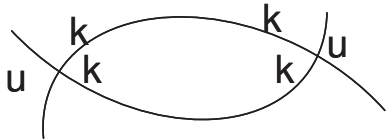, width=2in} 
  \caption{ The Ginzburg criterion requires the calculation of the above diagram.
    }
\label{fig1}
\end{figure} 

The idea of our calculations can be understood from the following example.
According to the Ginzburg criterion, deviations from the mean-field theory emerge if 
the diagram depicted in Fig. 1 diverges in the infra-red limit, {\it i.e.}
when incoming momenta and frequencies are low. 
The diagram reduces to the integral
$I=\int dk d\omega k^4/[\omega^2+2ic\omega k +b k^4]^2$. We substitute $\omega=\omega'-ick$ and shift the integration contour
over $\omega'$ onto the real axis. This gives $I=\int dk d\omega' k^4/[\omega'^2+c^2k^2+bk^4]^2$. The latter integral converges.

To calculate the magnetization we add to the action a weak magnetic 
field term, $-h_q iq\phi_s(q)/(4\pi)$, where $\phi_s(q)$ denotes 
a Fourier component of 
the spin Bose-field. The uniform field corresponds to the limit $q\rightarrow 0$. 
At $h=0$ the magnetization is a non-analytic function of $r$ but for any finite 
magnetic field the magnetization $m$ depends on $r$ analytically 
\cite{stattheor,foot}. 
Thus, the critical behavior of the magnetization at small negative $r$ 
can be determined 
with the following three steps \cite{stattheor} : 
1) calculate 
$m=\langle\partial_x\phi_s\rangle/(4\pi)$ at $r>0$ using the perturbation theory 
about the $\phi_s=0$ state; 
2) analytically continue $m(r)$ to negative $r$; 
3) take the limit $h\rightarrow 0$. At $r>0$ we get the standard result 
\cite{stattheor}

\begin{equation}
\label{7}
\langle\partial_x\phi_s(x=0)
\rangle
=C\bigg\{\frac{h}{4\pi}-\sum_{n=2}^{\infty} \frac{1}{(2n-1)!}
\Gamma_{2n}(\tilde r;q_1=\dots=q_{2n}=\omega_1=\dots=\omega_{2n}=0)
\langle\partial_x\phi_s(x=0)\rangle^{2n-1}\bigg\};
\end{equation}
\begin{equation}
\label{8}
C=\lim_{q\rightarrow 0}\lim_{\omega\rightarrow 0}q^2G=1/\tilde r; 
G=1/[a \omega^2+2i\tilde c\omega q + \tilde rq^2],
\end{equation}
where $\Gamma_{2n}$ are the proper vertices at zero momentum and frequency, 
and $a,\tilde c$ and $\tilde r$ denote the renormalized coefficients of 
the Green function $G$. The formulas for the proper vertices
contain integrals of the products of the expressions of the form $p_1p_2G(\omega,q)$, where 
$p_1,p_2=\omega,q$. 
It is convenient to change variables: $\omega'=\omega+i\tilde cq/a$. Then 
$G\rightarrow G'=1/[a\omega'^2+(\tilde c^2/a+\tilde r)q^2]$. At $\tilde r>0$ 
the integration path for integrals over $d\omega'$ can be deformed into 
the real axis. 
(One can show this, e.g., using the representation 
$G'=\int_0^\infty ds \exp\{-s[a\omega'^2+(\tilde c^2/a+\tilde r)q^2]\}$ 
in Feynman integrals.)
The resulting integrals have no infrared divergences at
$\tilde r= 0$. Expanding $G'$ in powers of $\tilde r$ one 
can see that those integrals 
are analytic functions of $\tilde r$ at small $\tilde r$. 
Thus, all coefficients of Eq. (\ref{7}) are finite at $\tilde r=0$ and 
can be analytically continued to negative $\tilde r$. 
Eq. (\ref{7}) has the structure of the mean-field equation with the free energy 
$F=-hm+\tilde r(4\pi m)^2/2+\sum \Gamma_{2n}(\tilde r)(4\pi m)^{2n}/(2n)!$. 
Depending on the coefficients it may describe 
both first and second order phase transitions. 
A second order transition is possible for positive $\Gamma_4(\tilde r=0)$. 
In this case
the average magnetization follows the mean-field laws:

\begin{equation}
\label{9}
m(\tilde r<0,h=0)\sim\sqrt{-\tilde r}; 
\frac{dm(h=0,\tilde r>0)}{dh}\sim\frac{1}{\tilde r}; 
m(\tilde r=0,h)\sim h^{1/3}.
\end{equation}

The magnetization correlation function does not have the standard mean-field form. 
This is not surprising since in Luttinger liquids the correlation length is infinite 
for any $r$ and not only in the critical point \cite{Giamarchi}. 
Using the perturbation theory about the state with $\partial_x\phi_s=4\pi m(r)$ 
we find the large-distance behavior of the correlation function

\begin{equation}
\label{10}
\langle (\partial_x\phi_s(x=0,t=0)- 4\pi m)
( \partial_x\phi_s (x=X,t=0)-4\pi m)\rangle=
-\partial_x^2\int \frac{dk d\omega}{4\pi^2} \exp(ikX) G(\omega,k)
\sim 1/X^2.
\end{equation}
The coefficient before $1/X^2$ is a non-analytic function of $\tilde r$ near the phase transition.

So far we did not verify that $\cos(l\phi_s)$ is an irrelevant operator. 
This can be easily done with the renormalization group. 
At each step we represent $\phi_s=4\pi mx+\phi_s^<+\phi_s^>$ 
as the sum of the average 
$4\pi mx$, slow fluctuating part 
$\phi_s^<$, and fast part $\phi_s^>=
\int_{-\infty}^\infty d\omega\int_{b\Lambda<|k|<\Lambda} dk 
\exp(i\omega \tau+ikx)\phi_s(\omega,k)$, 
where 
$\Lambda$ is the ultra-violet cutoff and $b\ll 1$. 
We integrate out the fast modes $\phi_s^>$ and make the rescaling 
$k\rightarrow k/b$, $\omega\rightarrow\omega/b$.  
One can easily find the renormalization 
of the cosine term in the zero-loop order. We see that 
$\cos(l\phi_s)\rightarrow b^{-2}\cos(4\pi l mx/b + l\phi_s^<)
\exp(-l^2\langle\phi_s^{>2}\rangle/2)$, where $\langle\phi_s^{>2}\rangle\sim
\int d\omega \int_{b\Lambda<|k|<\Lambda} dk G(\omega,k)
\sim -(\ln b)/\sqrt{\tilde c^2+a\kappa |\tilde r|}$. 
Thus, the scaling dimension of the operator 
$\cos(l\phi_s)$, $2-{\rm const}/\sqrt{\tilde c^2+a\kappa |\tilde r|}$,  
is negative for small $\tilde c\sim V$ and $\tilde r$. 
Hence, the operator is irrelevant near the phase transition. 


The above perturbative argument applies only if the coefficients $w_l$ before 
$\cos(l\phi_s)$ are small. This explains why our results are valid in the case 
of the Ising symmetry $Z_2\times U(1)$ but not for the Heisenberg symmetry group even 
though the bosonized action has the same structure for both symmetries. 
Indeed, in the Heisenberg case $w_1$ and $r$ are related and cannot be simultaneously 
small \cite{Giamarchi}. 

The equilibrium ferromagnetic transition is described by the action 
$L=(\partial_\tau\phi_s)^2/2+r(\partial_x\phi_s)^2/2 +b(\partial_x^2\phi_s)^4+
u(\partial_x\phi_s)^4$ 
\cite{1fm2}. After the substitution 
$m=\partial_x\phi_s/(4\pi)$ this action reduces to the free energy 
of the uniaxial 
ferroelectric \cite{lkh}. Interestingly, the same universality class emerges in 
a number of unrelated non-equilibrium classical problems with conserved order parameter \cite{sm, feldman} 
and in equilibrium quantum problems \cite{v}. 
Long range dipole forces reduce the critical dimension of the uniaxial 
ferroelectrics to 3. This is equivalent to 2 in the quantum problem. 
Non-equilibrium effects are known to suppress fluctuations in classical systems 
\cite{feldman,mukamel}. Our results show that the same tendency exists in quantum
itinerant systems with conserved $S_z$ where the mean-field behavior is possible in 1D. 
On the other hand, non-equilibrium systems whose order parameter does not conserve
exhibit typically the same critical behavior as corresponding equilibrium systems \cite{review}.



In conclusion, we have shown that the voltage bias modifies the critical behavior 
of one-dimensional itinerant electrons near the ferromagnetic transition. 
The exact critical exponents are given by the mean field theory.


\begin{thebibliography} {150}
  
\bibitem{zrs1}
R. G. Mani {\it et al.}, 
Nature {\bf 420}, 6464 (2002);
M. A. Zudov {\it et al.}, 
Phys. Rev. Lett. {\bf 90}, 
046807 (2003).

\bibitem{zrs2}
J. Alicea {\it et al.}, 
cond-mat/0408661.

\bibitem{Aranson} I. S. Aranson {\it et al.}, 
Phys. Rev. Lett. {\bf 92}, 234301 (2004).

\bibitem{dds} 
B. Schmittmann and R. K. P. Zia, 
{\it Statistical Mechanics of Driven Diffusive Systems. 
Vol. 17 of Phase Transitions and Critical Phenomena},  
eds. C. Domb and J. L. Lebowitz 
(Academic Press, London, 1995). 

\bibitem{coldatoms}
M. R. Andrews {\it et al.}, 
Science {\bf 273}, 84 (1996).

\bibitem{apdhl} 
A. Polkovnikov {\it et al.}, 
cond-mat/0412497.

\bibitem{it} T. R. Kirkpatrick {\it et al.},
Phys. Rev. B {\bf 53}, 14364 (1996).

\bibitem{it1} D. Belitz {\it et al.}, 
Phys. Rev. Lett. {\bf 82}, 4707 (1999).

\bibitem{it2} D. Belitz {\it et al.}, 
Phys. Rev. B {\bf 63}, 174427 (2001).

\bibitem {it3} Ar. Abanov {\it et al.},
Phys. Rev. Lett. {\bf 93}, 255702 (2004).

\bibitem{LM}
E. H. Lieb {\it et al.}, 
Phys. Rev. B {\bf 125}, 164 (1962).

\bibitem{sim}
S. Daul {\it et al.},
Phys. Rev. B {\bf 58}, 2635 (1998).

\bibitem{fm07}
K. J. Thomas {\it et al.}, Phys. Rev. Lett. {\bf 77}, 135 (1996).

\bibitem{07}
D. J. Reilly {\it et al.}, Phys. Rev. Lett. {\bf 89}, 246801 (2002).

\bibitem{1fm1}
L. Bartosch {\it et al.}, 
Phys. Rev. B {\bf 67}, 092403 (2003)

\bibitem{1fm2}
K. Yang, Phys. Rev. Lett. {\bf 93}, 066401 (2004).

\bibitem{lkh}
A. I. Larkin {\it et al.},
JETP {\bf 29}, 1123 (1969).

\bibitem{sm}
B. Schmittmann, Europhys. Lett. {\bf 24}, 109 (1993).

\bibitem{feldman}
D. E. Feldman, JETP Lett. {\bf 61}, 953 (1995). 

\bibitem{leads} D. L. Maslov {\it et al.},
Phys. Rev. B {\bf 52},
  R5539 (1995); V. V. Ponomarenko, {\it ibid.} {\bf 52}, R8666 (1995);
  I. Safi {\it et al.},
{\it ibid.} {\bf 52}, R17040 (1995).
  
\bibitem{FG} D. E. Feldman {\it et al.}
Phys. Rev. B {\bf 67}, 115337
  (2003).
  
\bibitem{TSDG} B. Trauzettel {\it et al.}, 
  Phys. Rev. Lett. {\bf 92}, 226405 (2004); cond-mat/0409320.

\bibitem{FSV} D. E. Feldman {\it et al.}, 
Phys. Rev. Lett. {\bf 94}, 186809 (2005).

\bibitem{Keldysh} J. Rammer {\it et al.},
Rev. Mod. Phys. {\bf 58},
  323 (1986).

\bibitem{Giamarchi} 
T. Giamarchi, {\it Quantum Physics in One Dimension} (Claredon Press, Oxford, 2004).

\bibitem{Haldane}
F. D. M. Haldane, J. Phys. C {\bf 14}, 2585 (1981).

\bibitem{stattheor}
J. Zinn-Justin, {\it Quantum Field Theory and Critical Phenomena} 
(Claredon Press, Oxford, 2002).
  
\bibitem{foot} We assume that there are no phase transitions at small non-zero $h$ 
in the vicinity of $r=0$.

\bibitem{v} A. Vishwanath {\it et al.}, 
cond-mat/0311085;
E. Fradkin {\it et al.}, 
cond-mat/0311353.

\bibitem{mukamel} M. R. Evans {\it et al.}, 
Phys. Rev. Lett. {\bf 74}, 208 (1995).
  
\bibitem{review}
G. Odor, Rev. Mod. Phys. {\bf 76}, 663 (2004).

\end{thebibliography}
\end{document}